\documentstyle[12pt,epsf,epsfig]{article}
\def\s{{\cal S}}

\def\tr{{\rm tr}}

\def\gappeq{\mathrel{\rlap {\raise.5ex\hbox{$>$}}
{\lower.5ex\hbox{$\sim$}}}}
\def\permil{$\%\raise.20ex\hbox{$_0$}}
\def\lappeq{\mathrel{\rlap{\raise.5ex\hbox{$<$}}
{\lower.5ex\hbox{$\sim$}}}}
\begin{document}
\topmargin -1.0cm
\oddsidemargin -0.4cm
\evensidemargin -0.4cm
\pagestyle{empty}
\begin{flushright}
UG-FT-127/01\\
\end{flushright}
\vspace*{5mm}
\begin{center}

{\Large\bf TeV Strings and the 
Neutrino-Nucleon Cross Section

at Ultra-high Energies
}\\

\vspace{2cm}
{\sc F. Cornet, J.~I. Illana and M. Masip}\\
\vspace{.8cm}
{\it Departamento de F\'\i sica Te\'orica y del Cosmos}\\
{\it Universidad de Granada}\\
{\it E-18071 Granada, Spain}\\

\end{center}
\vspace{1.4cm}
\begin{abstract}

In scenarios with the fundamental unification scale at the
TeV one expects string excitations of the standard model
fields at accessible energies.
We study the neutrino-nucleon cross section in these
models. We show that duality of the scattering amplitude
forces the existence of a tower of 
massive leptoquarks that mediate the process in the
$s$--channel. Using the narrow-width approximation 
we find a sum rule for the production rate 
of resonances with different spin at each mass level. 
We show that these contributions can increase 
substantially the standard model neutrino-nucleon 
cross section, although seem insufficient in order to explain
the cosmic ray events above the GZK cutoff energy.

\end{abstract}

\vfill

\eject

\pagestyle{empty}
\setcounter{page}{1}
\setcounter{footnote}{0}
\pagestyle{plain}


\section{Introduction}

Extensions of the Standard Model (SM) with extra dimensions 
offer new ways to accommodate the hierarchies observed 
in particle physics \cite{Antoniadis:1990ew}. A very attractive
possibility would be to bring the scale of unification 
with gravity from $M_{\rm Planck}\approx 10^{19}$ GeV down to the 
electroweak scale $M_{\rm EW}\approx 1$ TeV. This 
could result if gravity propagates along 
a (4+$n$)--dimensional (--d) flat space with $n$ compact 
submillimeter dimensions (ADD model \cite{Arkani-Hamed:1998rs}) 
or along a (4+1)--d slice of anti-deSitter space with a warp
factor in the metric (RS model \cite{Randall:1999ee}).
These higher dimensional field theories, however, must be 
considered effective low-energy limits only valid below the
mass scale of a more fundamental theory. And nowadays,
only string theory \cite{pol98} provides a 
consistent framework for the unification of gravity with the 
standard model. 

At energies where 
the effects of a higher dimensional graviton are 
unsuppressed one expects the presence of its string Regge (SR)
excitations giving an effect of the same size.
This will be necessary in order to avoid the pathologies 
of spin-2 field theories. In string theory the massless
graviton comes as the zero mode of a closed string, whereas
the gauge bosons are the lightest modes of an open string. 
Now, as emphasized in \cite{Accomando:2000sj,Cullen:2000ef}, 
the exchange amplitude of a closed 
string has an order $g^2$ suppression versus the exchange of an 
open string. 
In consequence, processes that receive
sizeable contributions from SR and Kaluza-Klein excitations of the 
graviton and also from SR excitations of the gauge bosons will 
be dominated by the second ones.
This seems to be a generic feature in models of higher 
dimensional gravity embedded in a weakly-coupled string theory. 

In this paper we explore some phenomenological consequences
of the theories with unification at the TeV. In particular,
we focus on the neutrino-nucleon cross section. Our interest
is based on the possibility that new neutrino physics could
explain the cosmic ray events above the GZK cutoff energy 
(see \cite{Stecker:2001ek} and references therein).
We discuss a genuine string effect, the presence of leptoquarks
that mediate the process in the $s$-- and/or the $u$--channel.
This fact is a generic consequence required
by the duality of the scattering amplitudes. 
The leptoquarks appear at the massive SR level
even in string models where the only massless modes are
the SM fields (string models with the SM gauge symmetry).
The impact of leptoquarks on the neutrino-nucleon
cross section at ultra-high energies was first 
studied in \cite{Doncheski:1997it}, whereas in 
\cite{Domokos:1999ry} they are proposed in a framework
of strongly interacting neutrinos. 

\section{The $\nu_L u_L \rightarrow \nu_L u_L$ string amplitude}

Cullen, Perelstein and Peskin build in Ref.~\cite{Cullen:2000ef} 
a TeV-string model for QED.
It contains electrons and photons at low energies and massive 
SR excitations above the string scale. These excitations 
give corrections to QED processes that can be easily 
calculated. In this section we generalize their results in order to
obtain the $\nu_L u_L \rightarrow \nu_L u_L$ string amplitude.

The model results from a simple embedding of the SM 
interactions into Type IIB string theory. It is assumed that 
the 10--d space of the
theory has 6 dimensions compactified on a torus with 
common periodicity $2\pi R$ (the case with 6 extra dimensions 
tends to alleviate the graviphoton problem \cite{Atwood:2001au}), 
and that $N$ coincident D3-branes (4--d hypersurfaces
where open strings may end) are stretched out in the 4 extended
dimensions. We also assume that 
the extra symmetry of the massless string modes 
can be eliminated by an appropriate orbifold projection,
resulting an acceptable model with (at least) the SM
fields. The parameters of this theory would be the
string scale $M_S=\alpha'^{-1/2}\approx 1$ TeV and the 
dimensionless gauge coupling contant $g$, 
unified at $M_S$. Proposals for splitting these couplings
can be found in \cite{Ibanez:2000pw}. For more general 
D-brane models see \cite{Antoniadis:2000jv} and references therein.

A tree-level amplitude of open string states 
on a D-brane is given 
\cite{Cullen:2000ef,Hashimoto:1996kf} as a sum of ordered 
amplitudes multiplied by 
Chan-Paton traces. For the process under study we have
\begin{eqnarray}
{\cal A}(1,2,3,4)&=&
g^2 \cdot \s(s,t) \cdot F^{1243}(s,t,u) \cdot 
\tr [t^1t^2t^4t^3+t^3t^4t^2t^1] \nonumber \\
&+& g^2 \cdot \s(s,u) \cdot F^{1234}(s,u,t) \cdot 
\tr [t^1t^2t^3t^4+t^4t^3t^2t^1] \nonumber \\
&+& g^2 \cdot \s(t,u) \cdot F^{1324}(t,u,s) \cdot 
\tr [t^1t^3t^2t^4+t^4t^2t^3t^1] \, .
\label{string0}
\end{eqnarray}
In this expression,
\begin{equation}
\s(s,t)={\Gamma(1-\alpha' s) \Gamma(1-\alpha' t)\over
\Gamma(1-\alpha' s-\alpha' t)}
\end{equation}
is basically the Veneziano amplitude \cite{Veneziano:1968yb}, 
$(1,2,3,4)$ label $(\nu^{in}_L, u^{in}_L, \nu^{out}_L, u^{out}_L)$,
the Chan-Paton factors $t^a$ are representation
matrices of $U(N)$, and 
$F^{abcd}(s,t,u)$ is a factor depending
on the vertex operators for the external states and
their ordering.
In our case all the vertex operators will correspond to (massless) Weyl
spinors of helicity (directed inward) $+$ or $-$, giving
\begin{eqnarray}
F^{-++-}(s,t,u)&=&-4{t\over s}\;; \nonumber \\
F^{-+-+}(s,t,u)&=&-4{u^2\over st}
=4 \left( {u\over s}+{u\over t}\right) \;;\nonumber \\
F^{--++}(s,t,u)&=&-4{s\over t}\;.
\end{eqnarray}
We obtain
\begin{equation}
{\cal A}(\nu_L u_L \rightarrow \nu_L u_L)=
-4 g^2 \left[ {s\over t} \; \s(s,t) \; T_{1243} + 
{s\over u} \; \s(s,u) \; T_{1234} + 
{s^2\over tu} \; \s(t,u) \; T_{1324} \right] \, ,
\label{ampl0}
\end{equation}
with $T_{abcd}$ the Chan-Paton traces.

To understand the phenomenological consequences of this amplitude
let us start with the limit $s,t\rightarrow 0$. Since $\Gamma(1)=1$,
we have all the Veneziano factors $\s (0,0)=1$. The amplitude
expresses then the exchange of massless vector modes in the $t$-- and 
the $u$--channels. The former would correspond to the $Z$ gauge 
boson, whereas the field exchanged in the $u$--channel is in the
$({\bf \overline 3}, {\bf 1})$ and/or the $({\bf \overline 3}, {\bf 3})$
representations of 
$SU(3)_C\times SU(2)_L$ and has electric charge $Q=-2/3$. 
The $SU(2)_L$-singlet can be found in the ${\bf 10}$ 
of $SU(5)$ or the adjoint ${\bf 45}$ of $SO(10)$, whereas
the triplet is, for example, in the ${\bf 35}$ of $SU(5)$. 
We are interested, however, in models that reproduce the SM result at
low energies, with no massless leptoquarks. We obtain this limit 
if the Chan-Paton factors assigned to
$u_L$ and $\nu_L$ are such that
$T_{1243}-T_{1324}=-{1\over 10}$  and $T_{1234}=T_{1324}$,
where we have used $\sin^2\theta_W=3/8$. 
In terms of $T_{1234}=-a/10$ the amplitude becomes
\begin{equation}
{\cal A}(\nu_L u_L \rightarrow \nu_L u_L)=
{2\over 5} g^2 \left[\; {s\over t} \left[ \left( 1+a \right) \; 
\s(s,t) - a\; \s(t,u) \right] 
+ {s\over u} \left[ a\; \s(s,u) - 
a\; \s(t,u) \right] \;\right] \, .
\label{ampl1}
\end{equation}
At low $s$ this amplitude is 
${\cal A}_0\approx (2/5) g^2 {s/ t}$ and corresponds to the
exchange of a $Z$ boson in the $t$--channel.
The $Z$ is then a massless SR mode that will 
acquire its mass $M_Z$ only through the Higgs 
mechanism. We shall neglect the corrections of order 
$M_Z^2/M_S^2$ that may affect the massive SR modes. 

As the energy increases the Veneziano factor $\s(s,t)$ gives a 
series of poles (at $1-\alpha' s=0,-1,-2,...$) and zeroes
(at $1-\alpha' s-\alpha' t=0,-1,-2,...$). It can be expressed as
\begin{equation}
\s(s,t)=\sum_{n=1}^{\infty} 
{\alpha' t+\alpha' s -1\over \alpha' t+ n - 1}\;
{\prod_{k=0}^{n-1}\; (\alpha' t + k)\over 
(\alpha' s -n)\; (n-1)!}\;.
\end{equation}
At $s=nM_S^2$ the amplitude will describe the exchange
of a collection of resonances with the same mass 
and different spin (see below). Away from 
the poles the interference of resonances at different
mass levels will produce the usual soft (Regge) behavior 
of the string in the ultraviolet. Obviously, these
resonances are not stable and at one loop will get an 
imaginary part in their propagator. When the total 
width of a resonance (which grows with its mass)
is similar to the mass difference with the 
resonance in the next level one cannot see resonances  
and interference effects dominate also at $s=nM_S^2$.

Let us first analyze the case with $a=0$ in Eq.~(\ref{ampl1}). 
The amplitude is just 
${\cal A}(\nu_L u_L \rightarrow \nu_L u_L)=
(2/ 5) g^2  (s/ t) \cdot \s(s,t)$.
Near the pole at $s=nM_S^2$ it is
\begin{equation}
{\cal A}_n\approx {2\over 5} g^2 \; {nM_S^4\over t}\; 
{(t/M_S^2)\; (t/M_S^2 +1)\cdot...\cdot 
( t/M_S^2 + n-1 ) \over 
(n-1)!\; (s - nM_S^2)}\;. 
\end{equation} 
This amplitude corresponds to the 
$s$--channel exchange of massive leptoquarks 
in the $({\bf 3}, {\bf 3})$ representation of 
$SU(3)_C\times SU(2)_L$ with 
electric charge $Q=2/3$. At each pole we have 
contributions of resonances with 
a common mass $\sqrt{n}M_S$ but different
spin, going from zero to the order of the 
residue $P_n(t)={\cal A}_n\cdot (s - nM_S^2)$. 
In this case the maximum
spin at the $n$ level is $J=n-1$.
To separate these contributions 
we first write the residue in terms of the
scattering angle $\theta$, with 
$t=-(nM_S^2/ 2)(1-\cos \theta)$.
Then we express $P_n(\theta)$ as a linear combination of
the $d$--functions (rotation matrix elements):
\begin{equation}
P_n(\theta)= {2\over 5} g^2 nM_S^2 \;
\sum_{J=0}^{n-1} \alpha^J_n\;
d^J_{0,0}(\theta)\;.
\end{equation}
The coefficient $\alpha^J_n$ gives the contribution
to our amplitude of a leptoquark $X^J_n$
of mass $nM_S^2$
and spin $J$. For example, at the first SR level
we find a scalar resonance with $\alpha^0_1=1$, 
at $s=2M_S^2$ there is a single vector resonance
with $\alpha^1_2=1$, whereas at 
$s=3M_S^2$ there are modes of spin 
$J=2$ ($\alpha^2_3=3/4$) and $J=0$ ($\alpha^0_3=1/4$).

The general case with $a\not= 0$ is completely analogous,
with resonant contributions from the terms 
proportional to $\s(s,t)$ and $\s(s,u)$.  
Taking $u=-(nM_S^2/ 2)(1+\cos \theta)$
and expressing again the residue in terms of
$d$--functions we find the same type of resonances
but with different $\alpha^J_n$ coefficients: 
$\alpha^0_1=1+2a$, $\alpha^1_2=1$, 
$\alpha^2_3=3(1+2a)/4$ and $\alpha^0_3=1(1+2a)/4$.  

\section{The $\nu N$ total cross section}

From the resonant amplitude 
$\nu_L u_L\rightarrow X^J_n \rightarrow \nu_L u_L$ we can now obtain
the partial width 
$\Gamma^J_n\equiv \Gamma ( X^J_n \rightarrow \nu_L u_L)$:
\begin{equation}
\Gamma^J_n={g^2\over 40\pi}\; {\sqrt{n} M_S\; 
| \alpha^J_n | \over 2J+1}\;.
\end{equation}
Notice that for a given spin $J$, the variation with $n$ of 
$\alpha^J_n$ gives the {\it running} of the coupling with
the energy. We obtain numerically that the coupling of
heavier resonances decreases like the power law 
$\alpha^J_n\approx 1/n$.

The partial width $\Gamma^J_n$ can be used to obtain the 
cross section 
$\sigma^J_n(\nu_L u_L)\equiv \sigma (\nu_L u_L\rightarrow X^J_n)$ 
in the narrow-width approximation:
\begin{equation}
\sigma^J_n(\nu_L u_L)={4\pi^2\;\Gamma^J_n\over \sqrt{n} M_S}\;(2J+1)\;
\delta (s-nM_S^2)\;.
\end{equation}
At each mass level $n$ there is a tower of resonances of integer 
spin $J$ from 0 to $n-1$. We find a sum rule for the production 
rate $\sigma_n(\nu_L u_L)\equiv \sum_J \sigma^J_n(\nu_L u_L)$ 
of any of these resonances:
\begin{eqnarray}
\sigma_n (\nu_L u_L)&=&\left\{ \begin{array}{ll} 
\displaystyle {2\over 5}\;{\pi g^2\over 4}\;(1+2a)\;
\delta (s-nM_S^2)&{\rm for}\;\;n\;\;{\rm odd}
\vspace{0.2truecm} \\
\displaystyle {2\over 5}\;{\pi g^2\over 4}\;
\delta (s-nM_S^2)&{\rm for}\;\;n\;\;{\rm even}\;.
\end{array} \right.
\label{cs1}
\end{eqnarray}
This is equivalent (for $a=0$) to the production rate 
of a single resonance of mass $\sqrt{n} M_S$ and coupling 
$(2/5) g^2$ \cite{Doncheski:1997it}. In our opinion this is a 
very interesting result. The coupling of heavier SR modes
decreases quadratically with the energy, but the number 
of modes (and the highest spin) 
at each mass level $n$ grows also quadratically making
$\sum_J \alpha^J_n$ a constant independent of $n$.

In the narrow-width approximation 
$\sigma(\nu_L u_L) \equiv \sum_{n,J} \sigma(\nu_L u_L\rightarrow 
X^J_n \rightarrow {\rm anything})$ is then 
$\sigma(\nu_L u_L)=\sum_n \sigma_n(\nu_L u_L)$. In this limit the 
cross section is proportional to a collection of delta functions 
and thus all interference effects are ignored. This is a good
approximation as far as the total width of a resonance
is smaller than the mass difference with the next resonance
of same spin. Although the coupling (and any partial width)
decreases with the mass, the total width of
heavier resonances will grow due to the larger number
of decay modes that are kinematically allowed. We estimate
that contributions to $\sigma(\nu_L u_L)$ from
modes beyond $n\cdot (g^2/4\pi) \approx 1$ 
are a continuum. In this regime any cross section goes
to zero exponentially at fixed angle ($s$ large, $t/s$ fixed)
and like a power law at small angles ($s$ large, $t$ fixed) 
\cite{pol98}. We neglect these contributions. We will keep only 
resonant contributions from levels 
$n<n_{cut}= 50$ and will find that our result depends very mildly on 
the actual value of $n_{cut}$.

To evaluate the total $\nu N$ cross section we shall also 
need the elastic amplitudes $\nu_L d_L$, $\nu_L u_R$, $\nu_L d_R$
and $\nu_L \overline q_{L(R)}$. 
${\cal A}(\nu_L d_L \rightarrow \nu_L d_L)$ takes the same
form as the amplitude in Eq.~(\ref{ampl1}) with the changes 
$(2/5,a)\rightarrow (-3/5,a')$. 
The massive resonances exchanged in the 
$s$--channel are now an admixture of an 
$SU(2)_L$ singlet and a triplet. The singlet contribution
is required in $n$-even levels, otherwise an $SU(2)_L$ gauge 
transformation would relate the parameters $\alpha^J_n$ 
obtained here with the ones deduced from  
Eq.~(\ref{ampl1}).
In $n$-odd mass levels gauge invariance could be obtained
with no singlets for $a'=-(2+a)/3$.
The cross section $\sigma_n(\nu_L d_L)$ can be read from
Eq.~(\ref{cs1}) just by changing $(2/5,a)\rightarrow (3/5,a')$.

The calculation of amplitudes and cross sections for
$\nu_L\! \stackrel{_{(\_)}\ }{q_R}$
are completely analogous. We obtain
\begin{eqnarray}
\sigma_n(\nu_L u_R) &=&\left\{ \begin{array}{ll} 
\displaystyle {2\over 5}\;{\pi g^2\over 2}\;b\;
\delta (s-nM_S^2)&{\rm for}\;\;n\;\;{\rm odd} 
\vspace{0.2truecm} \\
0 & {\rm for}\;\;n\;\;{\rm even}\;.
\end{array} \right.
\label{cs3}
\end{eqnarray}
The cross sections $\sigma_n(\nu_L d_R)$, 
$\sigma_n(\nu_L \overline u_R)$ and
$\sigma_n(\nu_L \overline d_R)$ coincide with the
expression in Eq.~(\ref{cs3}) with the changes 
$(2/5,b)\rightarrow (1/5,b')$, 
$(2/5,b)\rightarrow (2/5,a)$ and 
$(2/5,b)\rightarrow (3/5,a')$, respectively.
For the left-handed antiquarks, 
$\sigma_n(\nu_L \overline u_L)$ and
$\sigma_n(\nu_L \overline d_L)$ can be read from
Eq.~(\ref{cs1}) by changing 
$(2/5,a)\rightarrow (2/5,b)$ and 
$(2/5,a)\rightarrow (1/5,b')$, respectively.

Now the total neutrino-nucleon cross section due
to the exchange of SR excitations can be very easily
evaluated. In terms of parton distribution functions
$q(x,Q)$ ($q=q_{L,R},\overline q_{L,R}$) in a 
nucleon $(N\equiv (n+p)/2)$ and the 
fraction of longitudinal momentum $x$, it is
\begin{equation}
\sigma (\nu_L N )= \sum_{n=1}^{n_{cut}} \sum_{q}
{\tilde \sigma_n (\nu_L q)\over nM_S^2}\; x\; q(x,Q)\;,
\label{cs4}
\end{equation}
where $x=nM_S^2/s$, $Q^2=nM_S^2$ and 
$\tilde \sigma_n (\nu_L q)$ is the factor multiplying
the delta function in the cross section $\sigma_n (\nu_L q)$.

In Fig.~1 we plot the neutrino-nucleon cross section
at energies from $10^2$ to $10^{13}$ GeV for 
$M_S=0.5,2$ TeV. 
We have used
the CTEQ5 parton distributions in the DIS scheme 
\cite{Lai:2000wy} 
extended to $x < 10^{-5}$ with the methods in 
\cite{Gandhi:1996tf}. 
We include the SM cross section and 
plot the string corrections for 
$a=a'=b=b'$ equal 0 and 5 (notice that in the first case
there are no $s$--channel resonances mediating the 
$\nu_L q_R$ amplitude). We do not include the modes beyond 
$n_{cut}=50$, where we expect that the narrow width 
approximation is poor. We also find that around $80\%$
of the effect comes from the ten first SR modes.

\section{Conclusions}

Cosmic rays hit the nucleons in
the atmosphere with energies of up to $10^{12}$ GeV. If the
string scale is in the TeV range, these cosmic rays 
have the energy required to explore the fundamental 
theory and its interactions. In particular, ultra-high
energy neutrinos are interesting since they can travel
long distances without losing a significant fraction of 
energy. In addition, the SM interactions of a neutrino
are much weaker than those of a quark or a charged lepton,
which makes easier to see deviations due to new physics.

With this motivation we have analyzed the string
$\nu N$ cross section at energies much larger than
$M_S\approx 1$ TeV. We fix the arbitrary parameters
of the model (four Chan-Paton traces) imposing 
phenomenological constraints, namely,
the massless SR modes must account for the 
electroweak gauge bosons only. Then we find that
the massive SR modes include leptoquarks that 
mediate the process in the $s$--channel. 
The presence of massive leptoquarks 
is not a peculiarity of our toy model but a generic 
feature of any string model. The argument goes
as follows. If an amplitude is mediated
by a massless field in the $t$--channel, there
will be its higher-spin SR excitations 
mediating the process also in the $t$--channel. 
However, the only known way to make
sense of an amplitude mediated by 
(elementary) higher spin fields is {\it \`a la} 
Veneziano (open string) \cite{Veneziano:1968yb} 
or {\it \`a la} Virasoro (closed string) \cite{vir69}.
In the first case the amplitude has $(s,t)$ and/or 
$(t,u)$ duality (see {\it e.g.} Eq.~(\ref{ampl0})), 
whereas in the second case the amplitude has $(s,t,u)$
duality. In consequence, any amplitude with
$t$--channel poles will also have 
$s$-- and/or $u$--channel poles. Note 
that the $u$--channel poles of 
${\cal A}(\nu_L u_L)$ become $s$--channel 
poles of ${\cal A}(\nu_L \overline u_R)$.

We have found a very 
simple sum rule for the production rate of 
all the leptoquarks, with spin from 0 to $n\pm 1$, 
in the same mass level $n$. This made possible
the calculation of the total $\nu N$ cross 
section in the narrow-width approximation.
We obtain that the effect of these leptoquarks
is not just a correction of order 
$M^2_Z/M_S^2$ to the SM cross section, as one
would expect on dimensional grounds. 
SR excitations give a
contribution that can dominate for 
$M_S\approx 1$ TeV. This deviation (see Fig.~1) 
could explain the observation of
horizontal air showers in upcoming cosmic ray
experiments \cite{Tyler:2001gt}.
However, for the expected flux of ultra-high 
energy neutrinos it seems unlikely that the 
cosmic ray events observed above
the GZK limit correspond to the 
decay of resonances produced in $\nu N$ scattering. 
If the primordial
particle is a neutrino, the most promising
possibility would be $\nu \overline \nu$ scattering 
in the galactic halo with the 
the resonant production of a $Z$ boson 
\cite{Weiler:1999ny} or 
other massive field \cite{Davoudiasl:2000hv}
decaying into hadrons.

\section*{Acknowledgments} 

We thank N\'estor Armesto, 
Fred Olness, Ram\'on Pascual, Alex 
Pomarol, Hallsie Reno, Jos\'e Santiago and 
Angel Uranga for discussions. 
This work was supported by MCYT under contract FPA2000-1558,
by the Junta de Andaluc\'\i a under contract FQM-101, 
and by the European Union under contract HPRN-CT-2000-00149.

\newpage

\setlength{\unitlength}{1cm}
\begin{figure}[htb]
\begin{picture}(8,19.5)
\epsfxsize=16.cm
\put(0.0,2.5){\epsfbox{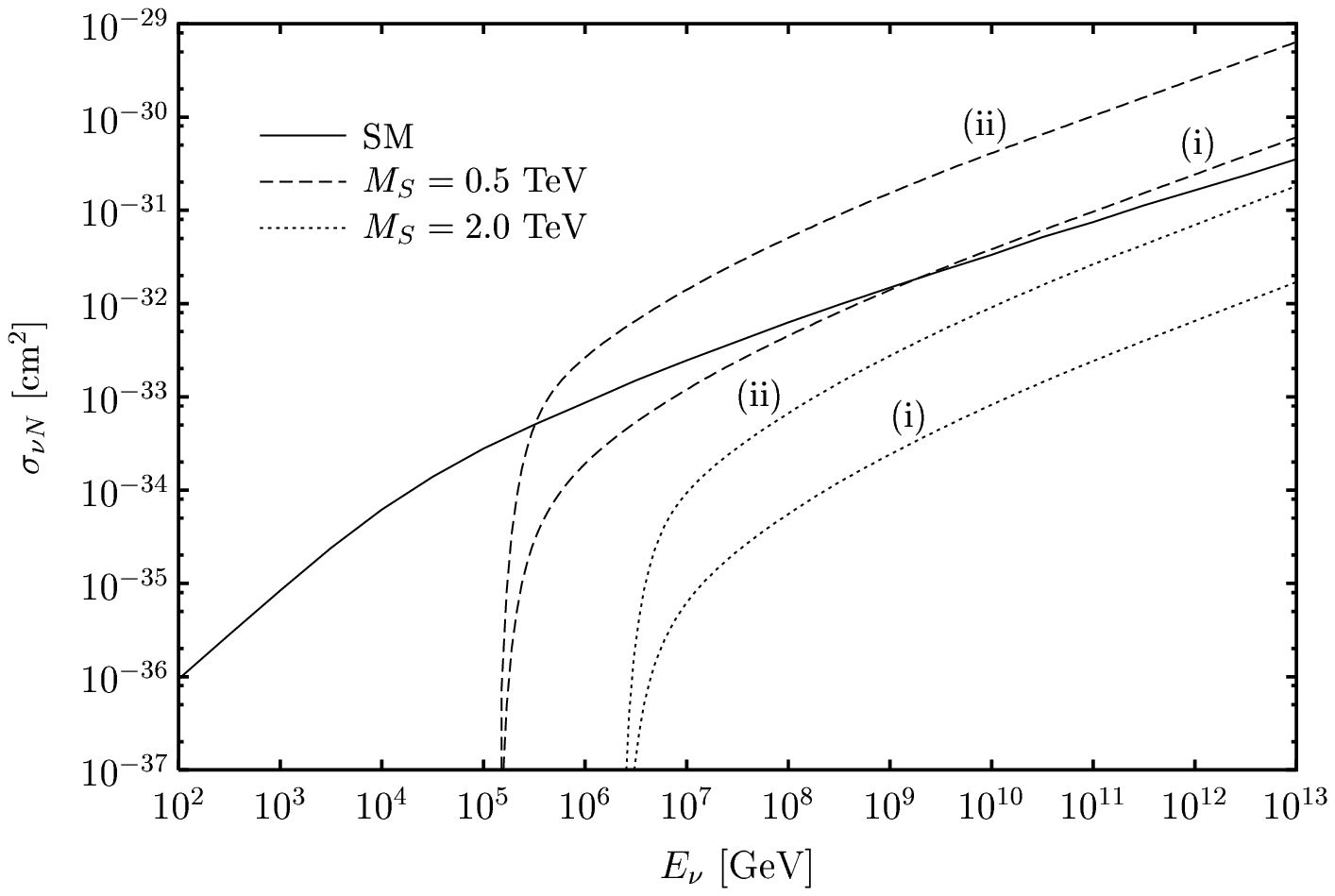}}
\end{picture}
\caption{
Neutrino-nucleon cross section versus the
incident neutrino energy $E_\nu$. The SM contribution
(solid) includes neutral and charged current interactions.
We plot the SR contribution for $M_S=0.5$ TeV (dashes)
and $M_S=2$ TeV (dots) for the cases (i)
$a=a'=b=b'=0$ and (ii) $a=a'=b=b'=5$.
\label{Figl}}
\end{figure}

\end{document}